\newif\ifcheckpagelimits
\checkpagelimitsfalse

\ifcheckpagelimits
 \documentclass[nofootinbib,prl,aps,twocolumn,showpacs,showkeys,%
 amsmath,amssymb,superscriptaddress,reprint,floatfix]{revtex4-1}
\else
 \documentclass[prl,aps,twocolumn,showpacs,showkeys,%
 amsmath,amssymb,superscriptaddress,reprint,floatfix]{revtex4-1}
\fi

\usepackage{amsfonts}
\usepackage{amsmath}
\usepackage{graphicx}
\usepackage{bm}
\usepackage{hyperref}
\usepackage{braket}
\begin{document}

\newcommand{\func}[1]{\mathrm{#1}}

\title{Ultrastrong coupling few-photon scattering theory}
\author{Tao Shi}
\affiliation{Max-Planck-Institut f\"ur Quantenoptik, Hans-Kopfermann-Stra{\ss}e 1, 85748 Garching, Germany}
\author{Yue Chang}
\affiliation{Max-Planck-Institut f\"ur Quantenoptik, Hans-Kopfermann-Stra{\ss}e 1, 85748 Garching, Germany}
\author{Juan Jos\'e Garc{\'\i}a-Ripoll}
\affiliation{Instituto de F{\'\i}sica Fundamental IFF-CSIC, Calle Serrano 113b, 28006 Madrid, Spain}
\date{\today}

\begin{abstract}
We study the scattering of photons by a two-level system ultrastrongly coupled to a one-dimensional waveguide. Using a combination of the polaron transformation with scattering theory we can compute the one-photon scattering properties of the qubit for a broad range of coupling strengths, estimating resonance frequencies, lineshapes and linewidths. We validate numerically and analytically the accuracy of this technique up to $\alpha=0.3$, close to the Toulouse point $\alpha=1/2$, where inelastic scattering becomes relevant. These methods model recent experiments with superconducting circuits [P. Forn-D{\'\i}az et al., Nat. Phys. (2016)].
\end{abstract}

\ifcheckpagelimits
\else
 \maketitle 
\fi

Waveguide quantum electrodynamics (QED) studies the interaction between propagating photons and quantum impurities in 1D environments. Reduced dimensionality empowers few-level systems with strong nonlinear features, and they become capable of fully reflecting individual photons\ \cite{astafiev10} or mediating a strong photon-photon interaction\ \cite{hoi13}. In order for this to occur, the impurity --a two-level system or qubit--- needs to be in the strong-coupling regime, where the spontaneous emission into the waveguide, $\Gamma$, dominates all other dissipation channels. This regime is achieved in experiments with superconducting circuits\ \cite{astafiev10,hoi13}, neutral atoms\ \cite{tiecke14,goban15} and quantum dots in photonic crystals\ \cite{arcari14}. In most experiments spontaneous emission is slower than the atom or photon oscillation frequencies, $\Gamma\ll \Delta,\omega$, allowing for a rotating-wave approximation (RWA) and theoretical predictions based on one- and few-photon wavefunctions\ \cite{shen05,shen07a}, input-output theory\ \cite{fan10,caneva15} and path integral formalism\ \cite{shi09,shi15}.

Superconducting circuits are waveguide QED systems where the qubit-photon coupling can match the qubit and photon energies, $\Gamma\sim \Delta,\omega$. This so called ultrastrong-coupling regime (USC) causes the breakdown of RWA predictions, the excitation of qubit-photon entangled ground states\ \cite{sanchez-burillo14}, extremely broadband interactions\ \cite{peropadre13}, and a phase transition into the localization regime\ \cite{leggett87}. The USC was first demonstrated in resonators\ \cite{niemczyk10,forn-diaz10}, where it admits an analytic description\ \cite{braak11}. More recently, the USC regime has been explored using propagating photons in microwave guides\ \cite{forn-diaz16} and studying the resonance spectrum of the qubit in the transmission line. This new generation of experiments opens a very challenging theoretical problem: the integration of USC in the waveguide QED framework, moving beyond the study of dissipation\ \cite{leggett87}, to photon-qubit scattering and interactions.

While this question has been addressed using numerical methods such as Matrix Product States or MPS\ \cite{peropadre13,diaz-camacho16}, the Numerical Renormalization Group\ \cite{bera16},  in this work we develop fully analytic predictions for the photon-qubit interaction, which are accurate for a broad range of the USC regime. Our starting point is the spin-boson model for a waveguide of length $L$
\ifcheckpagelimits
a a a a a a a a a a a a a a a a\else
\begin{equation}
  H = \frac{\Delta}{2}\sigma^z + \sum_k \omega_k a_k^\dagger a_k + \frac{1}{\sqrt{L}}\sum_k g_k \sigma^x(a_k + a_k^\dagger),
  \label{eq:spin-boson}
\end{equation}
\fi
We do not work with this Hamiltonian, but build a transformed one $H_\text{p}=U_\text{p}^\dagger H U_\text{p}$ using an optimized polaron transformation $U_\text{p}$ that eliminates most of the qubit-photon entanglement in the ground state\ \cite{diaz-camacho16}. The new Hamiltonian $H_\text{p}$ can be manipulated and combined with scattering theory\ \cite{shi15} to predict the dynamics of few-photon wavepackets. We show results for the superconducting Ohmic spin-boson model, where the spectral function is linear up to a cutoff $\omega_c$
\ifcheckpagelimits
a a a a a a a a a a a a a a a a\else
\begin{equation}
  J(\omega) =\frac{2\pi}{L} \sum_k |g_k|^2
  \delta(\omega-\omega_k)\simeq \pi\alpha \omega^1 e^{-\omega/\omega_c}.
  \label{eq:Ohmic}
\end{equation}
\fi
Interestingly, we recover cutoff independent predictions for the resonance and linewidth of elastic single-photon scattering in the USC regime. These results are validated with analytics at the Toulouse point\ \cite{guinea85,toulouse69} at $\alpha=1/2$ and also with moderate-size matrix-product state (MPS) numerical simulations of the qubit spontaneous emission. Both methods attest the qualitative $(\alpha\in[0.3,0.5])$ and even quantitative $(\alpha\in[0,0.3])$ accuracy of our techniques in modeling new and state-of-the-art experiments such as the single-photon scattering with tuneable coupling qubits by Forn-D{\'i}az et al.\ \cite{forn-diaz16}. This work opens the door to studying multi-photon scattering in more complex experiments with transmons or $\Lambda$-level schemes, or the development of accurate models for photon mediated interactions in open waveguides, which would have important applications in the quantum simulations of Ising-like Hamiltonians\ \cite{kurcz14} and annealing.

\paragraph{Model setup.---}%
Our starting point is the Hamiltonian\ \eqref{eq:spin-boson} that models the interaction between a two-level system and a photonic waveguide with periodic boundary conditions. The Pauli matrices $\sigma^{x,z}$ are defined in the qubit basis $\ket{e}$ and $\ket{g}$ for excited and ground states. The qubit couples to photons with momenta $k$, with anihilation (creation) operators $a_k$ $(a_k^{\dagger })$. We will conduct analytic calculations with a linear dispersion $\omega_k=c|k|$ and couplings $g_k=\sqrt{\pi \alpha c \omega_k/2}e^{-\omega_k/2\omega_c}$ that reproduce the Ohmic spectral function \eqref{eq:Ohmic}, and numerics with a hard cut-off $\omega_k = \omega_c\sqrt{[1-\cos(k)]/2}$, couplings $g_k=\sqrt{\pi \alpha c \omega_k/2}$ and $L$ equal to the number of modes.

\begin{figure}[t]
  \centering
  \includegraphics[width=0.9\linewidth]{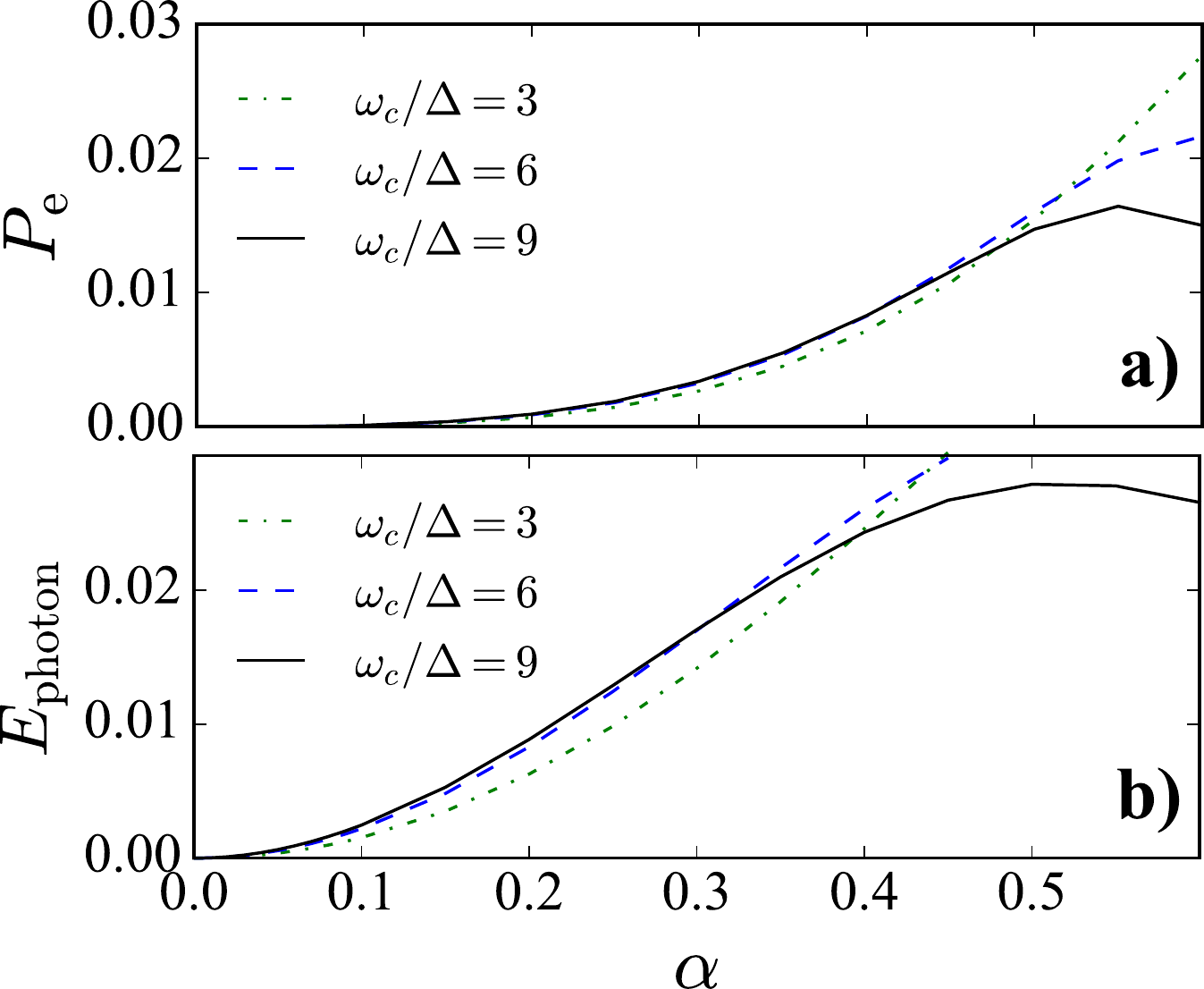}
  \caption{Ground state properties of the polaron Hamiltonian $H_\text{p}$ computed with MPS for different cutoffs. We plot (a) the excitation probability of the qubit $P_e=\braket{\sigma^z+1}/2$ and (b) the total energy of photons $E_\text{photon}=\sum_k\omega_k\braket{a^\dagger_ka_k}/\Delta$.}
  \label{fig:gs}
\end{figure}

Instead of\ \eqref{eq:spin-boson}, we implement approximations on the equivalent model $H_\text{p}=U_\text{p}^\dagger H U_\text{p}$  after a polaron transformation $U_\text{p}=\exp[-\sigma ^{x}\sum_kf_k(a_k^{\dagger }-a_k)/\sqrt{L}]$, which disentangles the bosonic and qubit states
\ifcheckpagelimits
a a a a a a a a a a a a a a a a\else
\begin{equation}
  H_\text{p} = \frac{\tilde\Delta}{2}\sigma^z O_{-\mathbf{f}}^\dagger O_{\mathbf{f}} + \sum_k\omega_ka^\dagger_ka_k
  +\sum_k \frac{G_k}{\sqrt{L}} \sigma^x(a_k + a_k^\dagger) + E_\text{p},
  \label{eq:polaron}
\end{equation}
\fi
The renormalized qubit energy $\tilde{\Delta}=\Delta e^{-2\sum_k|f_k|^2/L}$ appears with the operators $O_\mathbf{f}=\exp(2\sigma^x\sum_k f_k a_k/\sqrt{L})$ from normal ordering. The Silbey-Harris prescription\ \cite{silbey84,harris85} optimizes $f_k=g_k/(\omega_k+\tilde{\Delta}),$ reducing the effective coupling $G_k = \tilde{\Delta}f_k$, and making the ground state of $H_\text{p}$ as close to $\ket{g}\ket{0}$ as possible.

Despite the highly nonlocal term $\sigma^zO_{-\mathbf{f}}^\dagger O_\mathbf{f}$, it is possible to diagonalize $H_\text{p}$ using MPS ansatz\ \cite{ripoll06,verstraete08,orus14}, a variational estimate of the ground state wavefunction $\ket{\psi} = \sum_{s,\mathbf{n}}\mathrm{tr}\left[A^{s}_0A^{n_1}_1\cdots A^{n_N}_N\right] \ket{s,n_1\ldots n_N},$ where $A_i^x\in\mathbb{C}^{\xi\times\xi}$ are different matrices labeled by physical degrees of freedom: the qubit states, $s\in\{g,e\}$, or the photon occupation numbers $n_i$ of the associated momenta $k_i$. Numerical optimizations with the hard cut-off model $\omega_c/\Delta=3,6$ and $9$ show less than $2\%$ qubit excitation probability and a negligible amount of photons below the Toulouse point $\alpha=1/2$ [cf. Fig.\ \ref{fig:gs}]. Interestingly, most qubit-photon entanglement is removed by the polaron transformation and the MPS converges with a small bond dimension $\xi\ll 20$ (small matrices), and a small cut-off $n_i\leq 4$, significantly improving over earlier simulations with $H$\ \cite{peropadre13}.

\begin{figure}[t]
  \centering
  \includegraphics[width=0.9\linewidth]{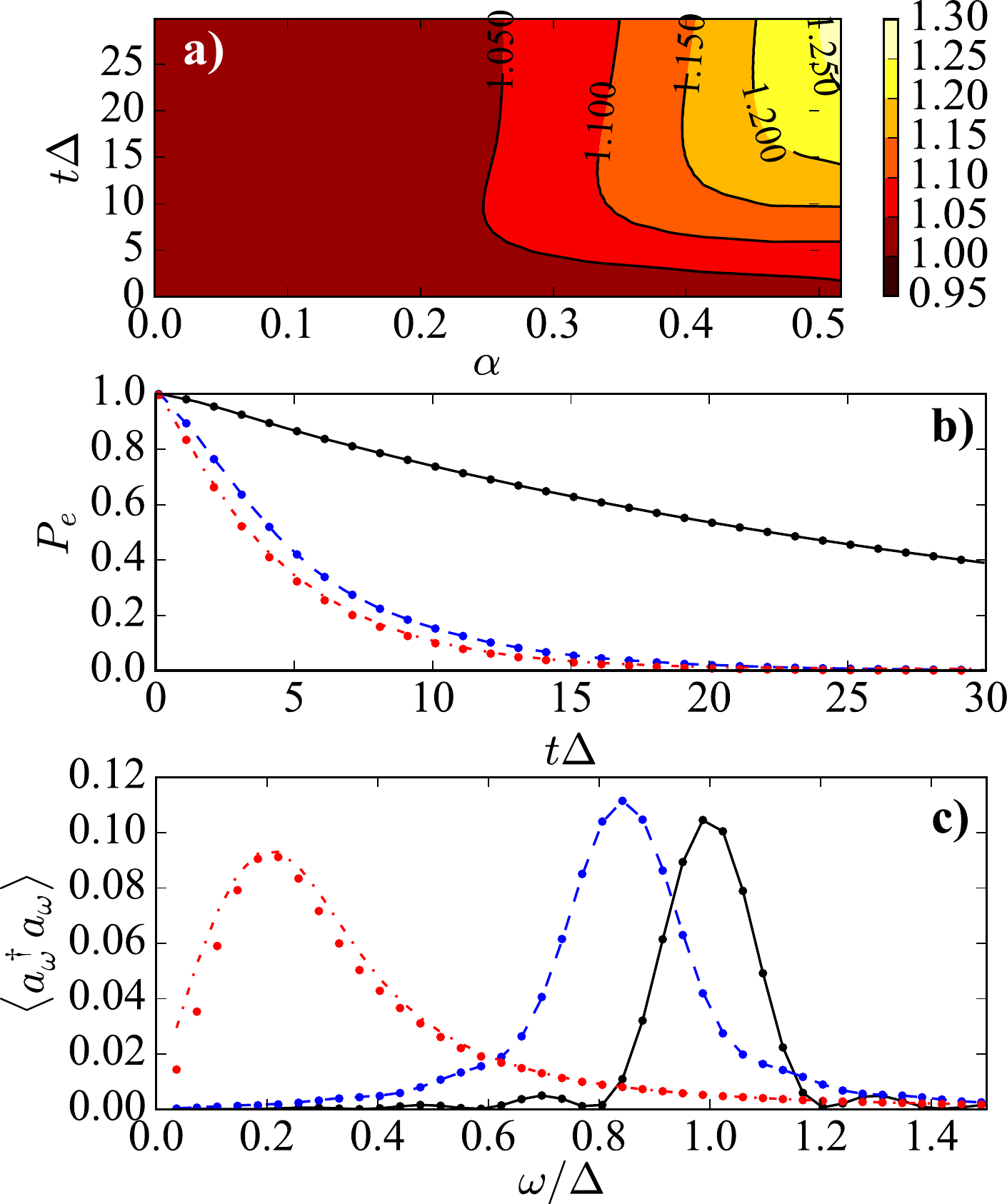}
  \caption{(Color online) Spontaneous emission of the two-level system in the polaron-transformed model $H_\text{p}$. (a) Total number of excitations $N=\sigma^+\sigma^-+\sum_k a^\dagger_k a_k$ starting from state $\ket{\uparrow}\ket{0}$. (b) Excitation probability $\braket{\sigma^+\sigma^-}$ as a function of time and (c) spectrum of emitted photons at $t\Delta=30$. Lines correspond to $\alpha=0.01$ (solid), $0.07$ (dashed) and $0.35$ (dash-dot), simulated with Hamiltonian\ \eqref{eq:polaron}. Thick dots represent the outcome from\ \eqref{eq:rwa-H} for similar values of $\alpha$.}
  \label{fig:emit}
\end{figure}

\begin{figure*}[t]
  \centering
\includegraphics[width=0.34\linewidth]{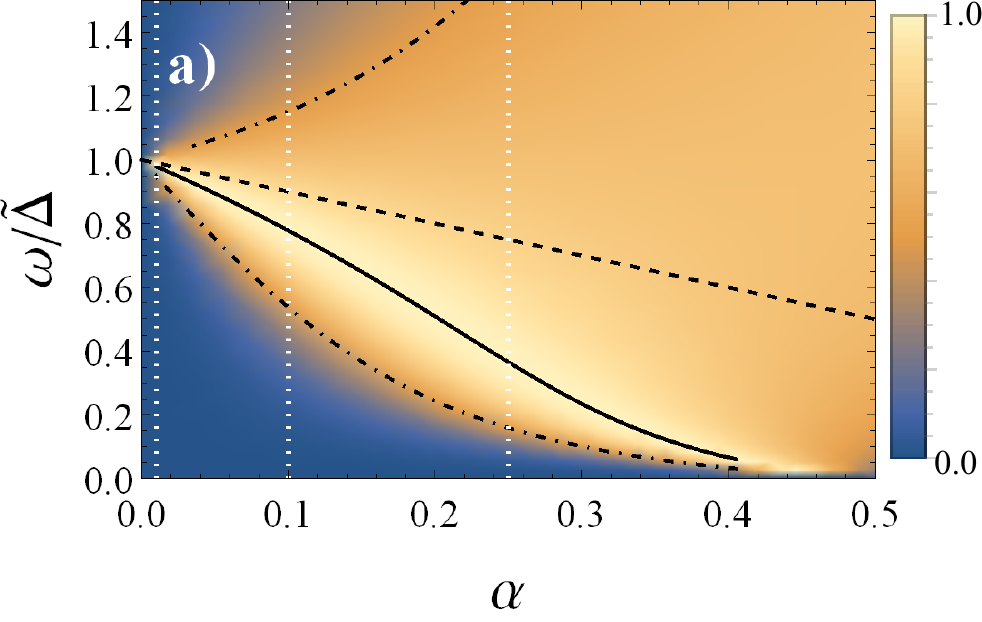}%
\includegraphics[width=0.32\linewidth]{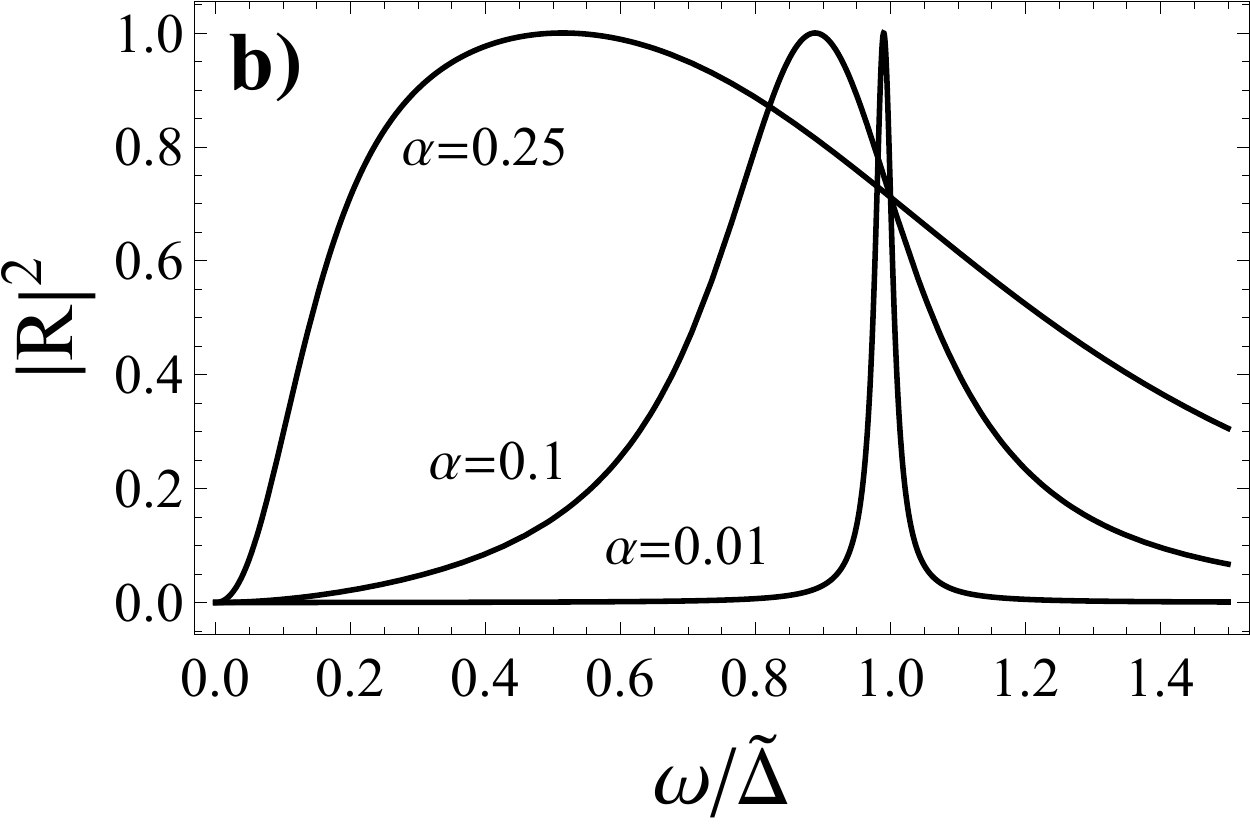}%
\includegraphics[width=0.32\linewidth]{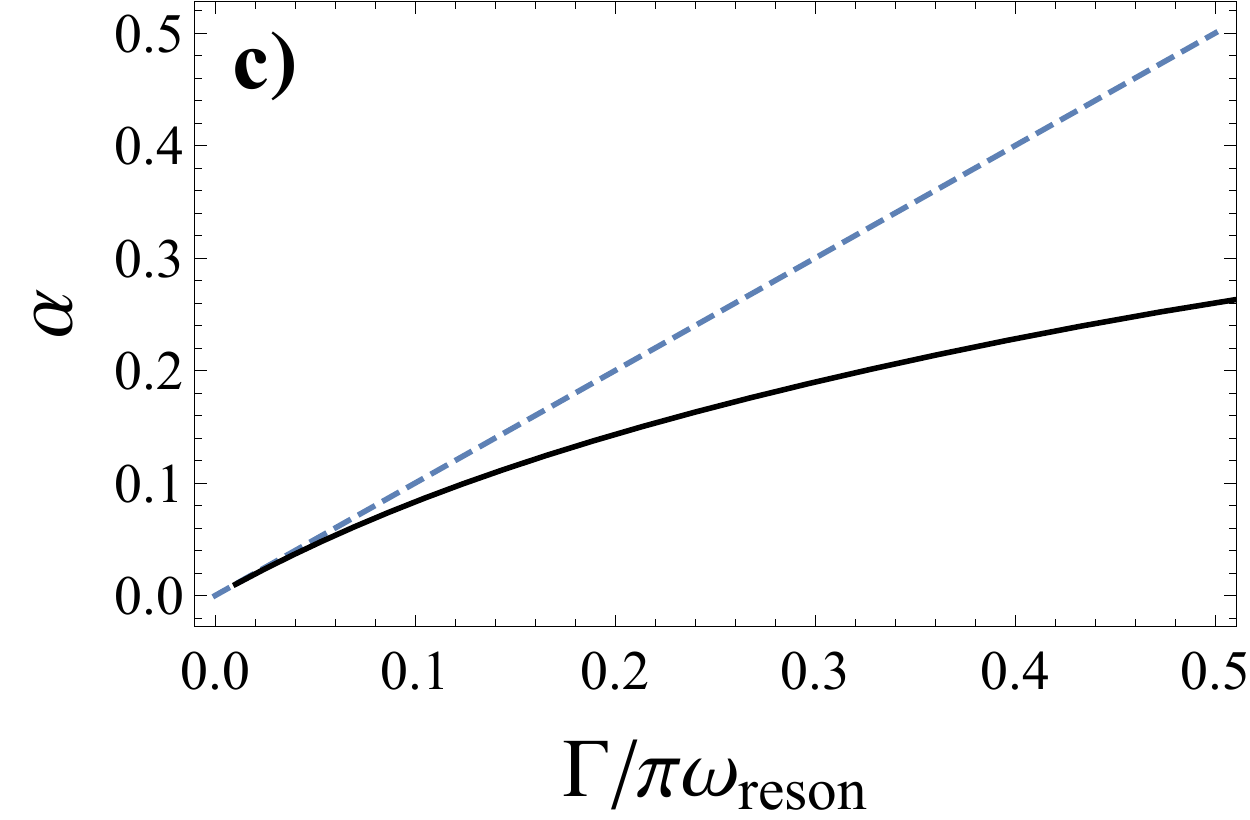}
\caption{(Color Online) (a) Reflection coefficient $R_k$ as a function of
the spin-boson coupling strength $\protect\alpha$ and the photon frequency.
In solid we plot the resonance frequency, $\protect\omega_{\text{reson}}$,
and in dashed-dot we show the half-height lines. (b) Three cuts of the above
plot show asymmetric lineshapes for increasing $\protect\alpha$. (c) The
ratio between the experimental linewidth and the resonance frequency is a
lower bound for $\protect\alpha$.}
\label{fig:RWA}
\end{figure*}

\paragraph{Excitation conserving polaron.---}%
Inspired by the simplicity of the ground state, we will now assume that the low-energy dynamics of $H_\text{p}$ admits also a simple description as quasiparticles on a close-to-vacuum state. For that we select the single-particle section Hamiltonian
\ifcheckpagelimits
a a a a a a a a a a a a a a a a a a a a a a a a a a a a a a a a\else
\begin{align}
  \label{eq:rwa-H}
H_\text{p}^{(1)} &=\frac{\tilde{\Delta}}{2}\sigma^z+V_\text{local}+\sum_{k,s=\pm}\omega_kA_{s,k}^{\dagger }A_{s,k} \\
&+\frac{2\sqrt{2}}{\sqrt{L}}\sum_k\tilde{\Delta}f_k(A_{+,k}^{\dagger}\sigma ^{-}+A_{+,k}\sigma ^{+}).\nonumber
\end{align}
\fi
This model introduces annihilation operators $A_{\pm ,k>0}=(a_k\pm a_{-k})/\sqrt{2}$ for the symmetric and anti-symmetric modes, and is restricted to work with a single excitation, $N=\sigma^+\sigma^- + \sum_k a^\dagger_ka_k=1$, which is sufficient for single photon emission and scattering. Note the spin-dependent potential, $V_\text{local}=-4\sigma^z\tilde{\Delta}\sum_{kp}f_kf_pA^\dagger_{+,p}A_{+,k}/L$, essential to capture the whole dynamics

We have compared both models using MPS simulations of a low-energy problem in which an excited qubit relaxes into a vacuum of photons, $\ket{\psi(0)}=\ket{e}\ket{0}$. As shown in Fig.\ \ref{fig:emit}a, the wavefunction at all times $\ket{\psi(t)}$ remains in the single excitation sector up to $\alpha\sim 0.3$. Moreover, Figs.\ \ref{fig:emit}b-c demonstrate an excellent agreement between the RWA\ \eqref{eq:rwa-H} and the Silbey-Harris Hamiltonian\ \eqref{eq:polaron}, not only in the qubit dynamics, but also in the emission spectrum. The quantitative disagreement is largely accounted for by (i) the additional dressing of localized photons by the qubit [cf. Fig.\ \ref{fig:gs}] and (ii) inleastic three- or more-photon contributions above $\alpha=0.35$\ \cite{goldstein13}.

\paragraph{Single-photon scattering estimates.---}%
The MPS suggest that we can work with $H_\text{p}^{(1)}$ up to $\alpha\simeq 0.3$. To get a feeling for this value, let us cut the waveguide to make a $\lambda/2$ resonator which is resonant with the qubit. We will find a qubit-cavity coupling $g_\text{cav}=\sqrt{\alpha}\Delta$. Thus, values of $\alpha=0.3$ correspond to $g_\text{cav}\simeq 0.55\Delta$ inside a cavity: a coupling so strong, that the bandwidth of photons is comparable to the qubit energy $\Gamma/\tilde{\Delta}\simeq 1$ [cf. Fig.\ \ref{fig:emit}c], the so called ultrastrong coupling regime.

Our goal is to analyse scattering in the USC regime, developing formulas that can be used to model experiments\ \cite{forn-diaz16}. We will apply scattering theory to $H_\text{p}^{(1)}$, focusing on the low power regime of at most one photon. The reflection and transmission coefficients
\ifcheckpagelimits
a a a a a a a a a a a a a a a a\else
\begin{equation}
  r_k = \frac{1}{2}(s_k-1),\; \mbox{and } t_k=\frac{1}{2}(s_k+1),\label{eq:r-and-t}
\end{equation}
\fi
are constructed from the chiral phase shift $|s_k|=1$ experienced by photons in the $A_{k,+}$ modes. These can be computed using scattering formalism\ \cite{shi15} or Lippmann-Schwinger theory
\ifcheckpagelimits
a a a a a a a a a a a a a a a a\else
\begin{equation}
s_k= \frac{(\omega_k-\tilde{\Delta})\tilde{\Delta}%
-(\omega_k+\tilde{\Delta})\Sigma ^{\ast }(\omega_k)}{%
(\omega_k-\tilde{\Delta})\tilde{\Delta}-(\omega_k+\tilde{%
\Delta})\Sigma (\omega_k)}.\label{phase}
\end{equation}
\fi
The self-energy $\Sigma (\omega)=\delta_L(\omega )-i\Gamma (\omega )/2$ contains a Lamb-shift
\ifcheckpagelimits
a a a a a a a a a a a a a a a a\else
\begin{equation}
\delta_L(\omega )=4\tilde{\Delta}^{2}\int_{0}^{\infty }\frac{dk}{2\pi }P\frac{f_k^{2}}{\omega -\omega_k},
\end{equation}
\fi
and a decay rate $\Gamma (\omega )=4\tilde{\Delta}^{2}f_{k_{0}}^{2}\left\vert \partial \omega_k/\partial k\right\vert _{k=k_{0}}^{-1}$ given by the solution $k_{0}\equiv k_{0}(\omega )$ of $\omega
_{k_0}=\omega $. The reflectivity $R_k=\left\vert r_k\right\vert ^{2}$ and transmissivity $T_k=\left\vert t_k\right\vert ^{2}$ determined by Eq.\ (\ref{phase}), satisfy $R+T=1$ and provide a concrete prediction for the lineshape of a single-photon scattering experiment, for all dispersion relations and frequency dependent couplings. When the dynamics of the spontaneous emission is slower than that of the qubit and photons, i.e., $|\Gamma |\ll \tilde{\Delta},\omega_k$, the Markov approximation reflects into a negligible potential $V_{\mathrm{local}}$ and Lamb-shift $\delta_L \sim 0$ and a uniform $\Gamma (\omega )\sim \Gamma $. We recover the usual formula
\ifcheckpagelimits
a a a a a a a a a a a a a a a a\else
\begin{equation}
r_k\simeq \frac{-i(\omega_k+\tilde{\Delta})\frac{\Gamma}{2} }{(\omega
_k-\tilde{\Delta})\tilde{\Delta}+i(\omega_k+\tilde{\Delta})\frac{%
\Gamma }{2}},
\end{equation}
\fi
predicting total reflection $R=1$ for resonant photons $\omega_k\rightarrow \tilde{\Delta}$, and displaying the usual Lorentzian profile from scattering experiments in the strong coupling regime with superconducting circuits\ \cite{astafiev10,hoi13} or quantum dots\ \cite{arcari14}. For USC, however, the self-energy and the local potential $V_{\mathrm{local}}$ induce significant distortions and asymmetries in the lineshapes, as expected from both earlier numerics\ \cite{peropadre13} and experiments\ \cite{forn-diaz16}.

\paragraph{Open transmission line.-}%
We particularize the predictions to the Ohmic coupling of a superconducting qubit with a transmission line. In the limit of large cut-off $\omega_c$, we may approximate $\tilde{\Delta} =\Delta \left({e\Delta }/{\omega_c}\right) ^{\alpha/(1-\alpha)}$, with varying prefactors depending on the details of the model. However, independent of the renormalization scheme, our scattering estimates lead to the  same self-energy%
\ifcheckpagelimits
a a a a a a a a a a a a a a a a\else
\begin{equation}
\Sigma (\omega )=\frac{2\tilde{\Delta}^{2}\alpha }{(\omega +\tilde{\Delta}%
)^{2}}\left[\omega \ln (\frac{\omega }{\tilde{\Delta}})-\omega -\tilde{\Delta}%
-i\pi \omega \right].  \label{selfenergy}
\end{equation}
\fi
This prediction does not involve the cut-off: this information is implicit in the value of $\tilde{\Delta}$, which the optimal transformation uses to regularize the couplings in both the infrarred and ultraviolet limits.

\begin{figure*}[t]
\includegraphics[width=0.32\linewidth]{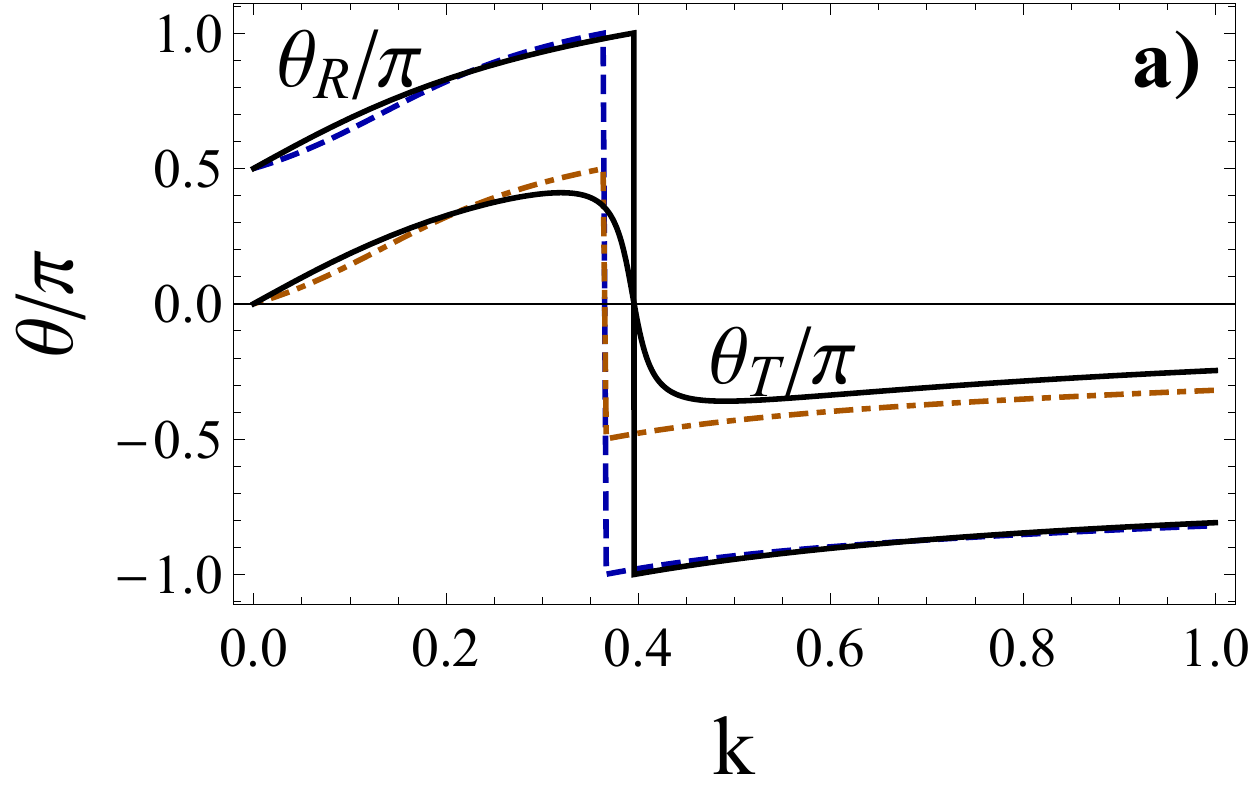}
\includegraphics[width=0.31\linewidth]{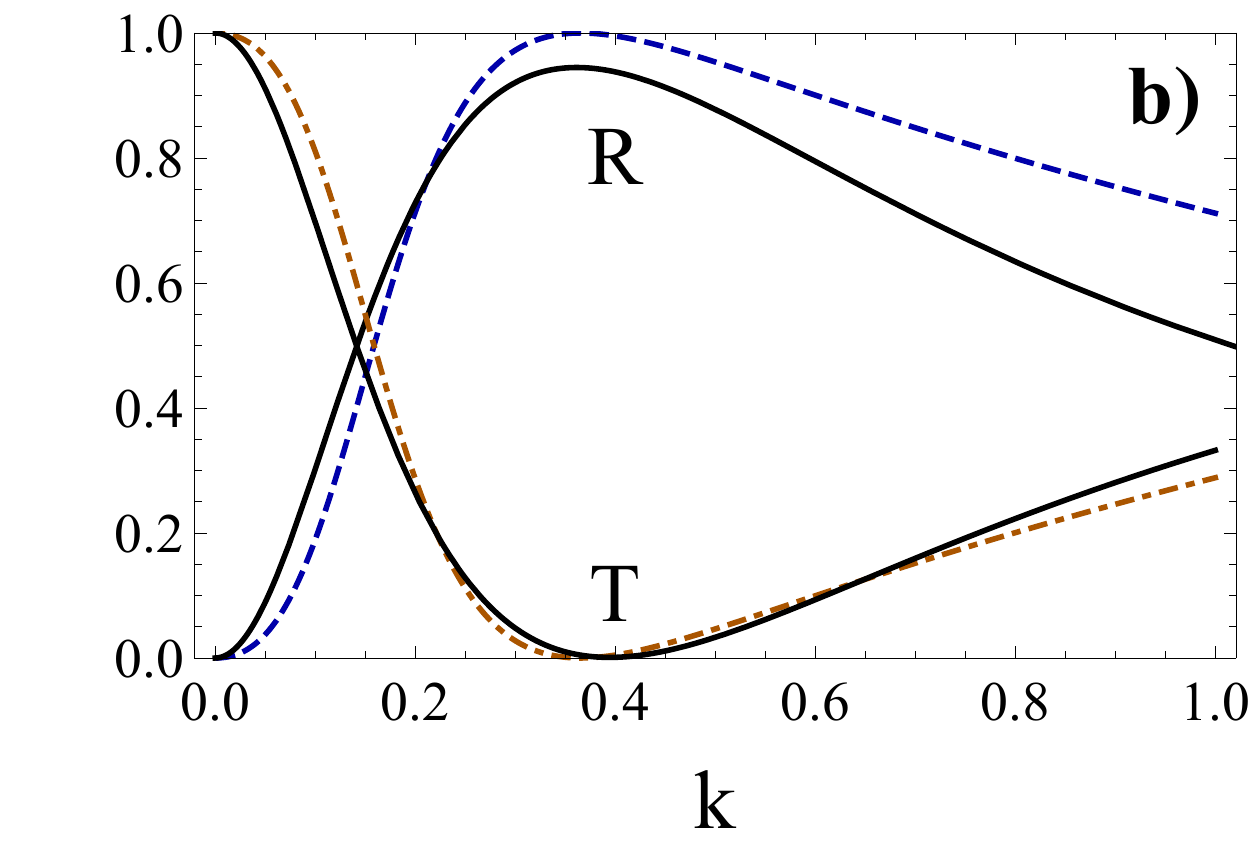}
\includegraphics[width=0.32\linewidth]{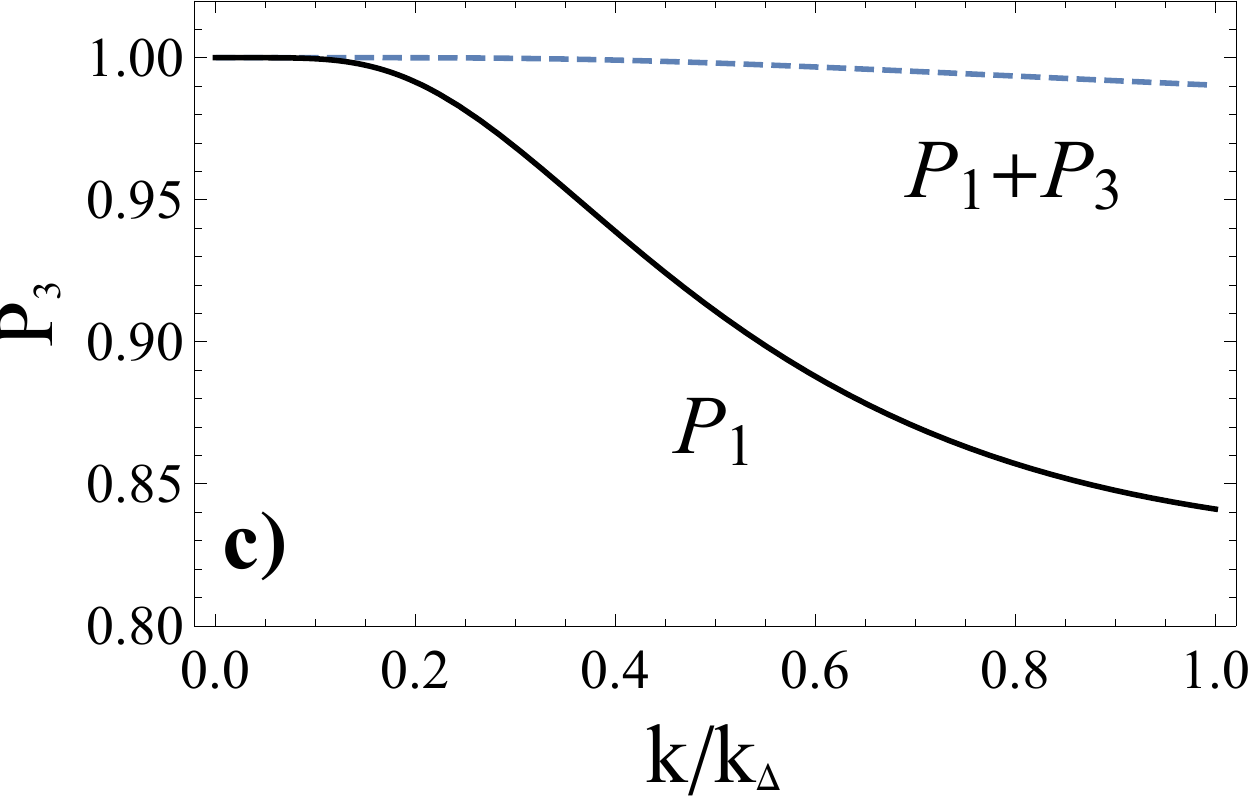}
\caption{(Color Online) Single-photon reflection and transmission coefficients as a function of the incident photon frequency at the Toulouse point $\alpha=1/2$, for $\omega_c=10^{8}\Delta$. We plot (a) the phases of the transmission and reflection amplitudes and (b) the reflection/transmission probabilities, both exact (solid) and with the polaron-RWA\ \eqref{phase} (dashed). (c) Probability to detect one and three out-going photons $P_1+P_3$ (dashed) and single-photon elastic scattering probability $P_1=R+T$ (solid).}
\label{fig:toulouse-1-photon}
\end{figure*}

The formula above has three important consequences: (i) The lineshape profiles are asymmetric for even moderate values of $\alpha\simeq 0.1$ [See Figs.\ \ref{fig:RWA}a-b]. (ii) As shown in Fig.\ \ref{fig:RWA}a, the scattering resonance $\omega_\text{reson}$, defined as the frequency of maximum reflection, does not necessarily match the value $\tilde{\Delta}$. This is due to a very large Lamb shift, of the same order of magnitude as the spontaneous emission rate itself. (ii) The linewidth $\Gamma$  depends on both $\alpha$ and $\omega$, and it is not possible to calibrate the interaction strength using the formula $\Gamma/\tilde{\Delta}\simeq \pi\alpha$, from the non-USC regime [cf. Fig.\ \ref{fig:RWA}c]. In other words, while it is true that we can distinguish the USC regime by the condition that $\Gamma/\pi$ be comparable to $\omega_\text{reson}$, a calibration of $\alpha$ demands the mathematical modelization of the line shapes.

\paragraph{Toulouse point, $\alpha=1/2$.---}%
While the above scattering formulas work for a broad range of couplings, $\alpha\in[0,0.3]$, it is interesting to study the source of deviations for very strong interactions, up to the phase transition into the Kondo regime. Fortunately, the spin-boson model with an Ohmic spectrum admits an analytical solution at the Toulouse point $\alpha=1/2$, which already has been used to study scattering properties\ \cite{goldstein13}.

The basic idea is to realize that at $\alpha=1/2$, working with a polaron displacement $f_k'=-g_k/\omega_k$ we can cancel completely the linear coupling terms, $G_k=0$, and map the spin and $A_+$ modes to the density fluctuations a fermionic bath, $A^\dagger_{q,+}=\sqrt{\frac{2\pi}{Lq}} \sum_k c_{k+q}^\dagger c_k$. The resulting model can be diagonalized, giving as ground state $\ket{\mathrm{GS}}=\ket{0}_{-}\ket{\text{FS}}_{f}$ a product of the vacuum $\ket{0}_{-}$ of the anti-symmetric modes and the Fermi sea $\ket{\text{FS}}_{f}$ of the new fermions. To compute the scattering matrix, we express the asymptotic state with one incoming photon $\ket{\phi _{\mathrm{in}}} =a_k^{\dagger }U_{\mathrm{p}}\ket{0}_{-}\ket{\text{FS}}_{f}$ using fermionic operators, and compute the  out-going  asymptotic state $\ket{\phi _{\mathrm{out}}} =\lim_{T\rightarrow \infty }e^{-iHT}\ket{\phi _{\mathrm{in}}}$ using the Green function approach in the infinite line limit $L\rightarrow \infty$.  The outgoing wavefunction contains both a single photon (elastic) component, as well as multiphoton (inelastic) contributions\ \cite{goldstein13}. The single-photon reflection and transmission\ \eqref{eq:r-and-t} derive from
\ifcheckpagelimits
a a a a a a a a a a a a a a a a a a a a a a a a a a a a a a a a\else
\begin{align}
  s_k =1+\frac{2iw}{1+2iw}[&2i(\func{arccot}2w+\arctan \frac{w}{1+2w^{2}})+\notag\\
 &+ \ln {w^{2}}/(1+w^{2})],
\end{align}
\fi%
with $w=\pi \Delta^{2}/(4\omega_ck)$. The value $s_k$ is no longer a phase and as a result, $P_1=|R_k|+|T_k|<1$. The true dynamics deviates from the polaron-RWA predictions because of multi-photon processes, which, as already shown in Ref.\ \cite{goldstein13}, are dominated by three-photon corrections $P_3=|\braket{0\vert \prod_{i=1}^{3}A_{+,p_{i}}^{\dagger }\vert \phi _{\mathrm{out}}}/\sqrt{3!}|^2$ with  momenta $p_1+p_2+p_3=k$. In Figs.\ \ref{fig:toulouse-1-photon}a-b we show the elastic transmission and reflection coefficients for both the exact Toulouse wavefunction and the polaron-RWA approximation. Considering that $\alpha=1/2$ is a very large interaction ($g_\text{cav}\simeq 0.71\Delta$ in the cavity), we find a very good qualitative and almost quantitative agreement between both methods in the elastic sector.

\paragraph{Summary and discussion.---}%
In this work we have derived analytical estimates of the lineshapes and resonance frequencies for single-photon scattering  in the USC regime, $\alpha\in[0,0.3]$ in the spin boson model ---or $g/\omega\in[0,55\%]$ if we would cut the same line to shape a cavity---. These estimates are supported by strong numerical evidence that the static and dynamic properties of the spin-boson model can be well approximated by a RWA version of the polaron transformation for this range of couplings. Our predictions apply to experiments with superconducting qubits in open transmission lines\ \cite{forn-diaz16} and represent an important milestone in the integration of the USC regime in waveguide QED theory. The techniques presented in this work can be immediately extended to other dispersion relations and coupling strengths, including, for instance, USC scattering in photonic bandgaps and cavity arrays with bound states\ \cite{sanchez-burillo14}.  It is also possible to account for radiative losses, heating and dephasing, depending on the qubit nature and its energy gap: in all cases the formulas generalize with the change $\Sigma(\omega) \to \Sigma(\omega) - i \Gamma_{\varphi}$, where $\Gamma_{\varphi}$ includes all additional dephasing sources.  We expect that the ideas put forward in this work will stimulate and simplify future experiments with superconducting circuits in the USC regime, as well as help in the development of a complete theory for USC scattering and effective interactions in multi-qubit setups.

\ifcheckpagelimits\else
\acknowledgments
JJGR acknowledges support from MINECO/FEDER Project FIS2015-70856-P and CAM PRICYT Project QUITEMAD+ S2013/ICE-2801. This work was funded by the European Union Integrated project Simulators and Interfaces with Quantum Systems (SIQS).
\bibliographystyle{apsrev4-1}
\end{document}